\documentclass[useAMS,usenatbib]{mn2e}

\usepackage{graphicx}
\usepackage{amssymb}
\usepackage{amsmath}
\usepackage{times}
\usepackage{color}
\usepackage{enumerate}

\newcommand{\pasp}{PASP}
\newcommand{\apjs}{ApJS}
\newcommand{\apj}{ApJ}
\newcommand{\apjl}{ApJL}

\newcommand{\aap}{A\&A}
\newcommand{\araa}{ARA\&A}
\newcommand{\aj}{AJ}
\newcommand{\mnras}{MNRAS}
\newcommand{\nat}{Nature}
\newcommand{\nar}{New Astron. Rev.}
\newcommand{\bain}{Bull. Astron. Inst. Netherlands}
 
\newcommand{\msun}{{\rm M}_\odot}

\begin{document}

\title[Ultraluminous X-ray sources in the Antennae galaxies]
{On the association of the ultraluminous X-ray sources in the Antennae galaxies with young stellar clusters 
\thanks{Based on observations obtained at the VLT/Melipal telescope, ESO, Paranal, Chile, in the framework of programmes 078.D-0766 and 080.D-0127.}}
\author[Poutanen et al.]
{Juri Poutanen,$^1$\thanks{E-mail: juri.poutanen@oulu.fi} 
Sergei Fabrika,$^2$  Azamat F.~Valeev,$^2$   Olga Sholukhova$^2$
and Jochen Greiner$^3$ \\
$^1$ Division of Astronomy, Department of Physics,  P.O. Box 3000, 90014 University of Oulu, Finland \\
$^2$ Special Astrophysical Observatory, Nizhnij Arkhyz 369167, Russia \\
$^3$ Max-Planck-Institut f\"ur Extraterrestrische Physik, Giessenbachstrasse 1, 85748 Garching, Germany}
 
\pagerange{\pageref{firstpage}--\pageref{lastpage}}
\pubyear{2013}
\date{Accepted 2013 March 16.  Received 2013 February 15; in original form 2012 October 2}
 
\maketitle
\label{firstpage}

\begin{abstract}
The nature of the ultra-luminous X-ray sources (ULXs) in the nearby galaxies is a matter of debates. 
One of the popular hypothesis associates them with  accretion at a sub-Eddington rate on to intermediate mass black holes. 
Another possibility is a stellar-mass black hole in a high-mass X-ray binary accreting at super-Eddington rates. 
In this paper we find a highly significant association between brightest 
X-ray sources in the Antennae galaxies and stellar clusters. 
On the other hand, we show that most of the  X-ray sources are located outside of these clusters. 
We study clusters associated 
with the ULXs using the ESO Very Large Telescope spectra and the {\it Hubble Space Telescope} data  
together with the theoretical evolutionary tracks and determine their ages to be below 6 Myr. 
This implies that the ULX progenitor masses certainly exceed 30 and for some objects are closer to 100 solar masses. 
We also estimate the ages of clusters situated close to the less luminous X-ray sources 
(with luminosity in the range $3\times 10^{38} \lesssim L_{\rm X} \lesssim 10^{39}$~erg s$^{-1}$) 
and find that most of them are younger than 10 Myr, because they are surrounded by strong H$\alpha$ emission. 
These findings are consistent with the idea that majority of ULXs are massive X-ray binaries 
that have been ejected in the process of formation of  stellar clusters by a few-body encounters and  
at the same time rules out the proposal that most of the ULXs are intermediate mass black holes.  
\end{abstract}
 

\section{Introduction}

Ultra-luminous X-ray sources (ULXs) are non-nuclear X-ray sources with luminosities exceeding 
an Eddington limit for a 10$\msun$ black hole \citep[see][ for a review]{2011NewAR..55..166F}. 
The most popular models for the ULXs involve either intermediate mass black holes (IMBH) with masses 
of $10^3 - 10^5\msun$ or stellar-mass black holes (StMBH) accreting at highly super-Eddington rates.
Both scenarios require massive donors in a close orbit (similar to SS 433, see \citealt{Fab04}), and 
the later model  also requires some collimation of radiation into our line of sight \citep{FabMesch2001,K01,PLF07} 
to explain the  objects up to possibly $10^{41}$ erg s$^{-1}$.

The X-ray luminosity functions in galaxies seems to have a universal powerlaw-like shape 
with a cut-off at luminosity of a few 10$^{40}$~erg s$^{-1}$ and a normalization 
proportional to the star formation rate \citep{Gilfanov03,Bregman06}. 
This implies that most of the ULXs are probably a high-luminosity extension of the  population of high-mass X-ray binaries.

%%%%
\begin{figure*}
\includegraphics[width=16cm]{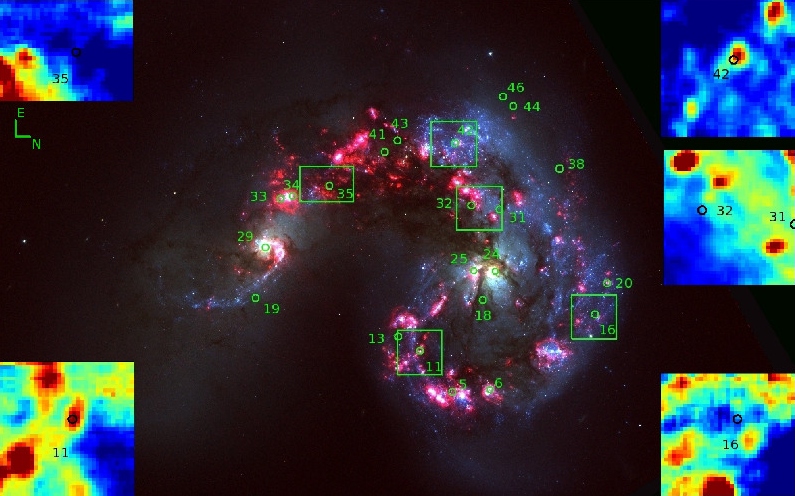}
\caption{The rgb image of Antennae produced from {\it HST}/ACS
images taken in FR656N (H$\alpha$, red), F435W (blue) and F550M (green) bands. 
NE bar length corresponds to 5\arcsec (107 pc/1\arcsec).
The VIMOS frames of the studied ULX regions together with Chandra positions (1\farcs0 radius) 
are indicated in green. The source numbers correspond to the catalogue of \citet{Zezas2002a}. 
The VIMOS continuum images have been extracted in the ACS band F550M and
the source positions are shown by black circles of 0\farcs3 in radius. 
}
\label{fig:wholemap}
\end{figure*}
%%%%

IMBHs originating from low-metallicity Population III stars can form binary systems due to tidal captures of single stars.
The expected frequency of such IMBHs is not high and does not agree with the ULX observed frequencies \citep{Kuranov07}. 
Moreover, these sources should be distributed throughout the galaxy, while 
most of the ULXs are associated with the star-forming regions \citep{Swartz2009}.
An IMBH produced in a runaway merging in a core of a young stellar cluster \citep{PortegiesZwart04,Freitag06} 
should stay within it,  because of the large IMBH mass.
   
On the other hand, stellar-mass black holes accreting at super-Eddington rates should 
be members of high-mass binaries. 
These objects are young and should be associated with clusters where are they born. 
However, they do not need to be in the cluster centres, instead they can be very effectively ejected from 
the clusters by few-body encounters at the initial stages of the cluster formation  \citep{Poveda67,Heggie75,Mikkola83b,Moeckel10,Kroupa10,Mapelli11}. 
Also the supernova explosions can eject massive binaries 
\citep{Zwicky57,Blaauw61,Shklovskii76,Woosley87,CordesChernoff98,vandenHeuvel2000}, 
but probably with lower velocities and  at a late stage of cluster evolution, when 
the massive stars had time to evolve.

Thus the distribution of the X-ray sources around stellar clusters as well as the cluster  age  
can give us a clue on the nature of ULXs and on the ejection mechanism. 
\citet{Kaaret04} have found that bright X-ray sources ($10^{36} < L_{\rm X} < 10^{39}$~erg s$^{-1}$) 
in three starburst galaxies are preferencially located near the star clusters,   but not within them. 
In a search for optical counterparts of 44 ULXs in 26 nearby galaxies, 
\citet{Ptak2006} have found that 28 of them have potential optical counterparts within $2\sigma$ error circles and 
10 of those have multiply counterparts. 
Their astrometric accuracy varied in the range 0\farcs3--1\farcs7 depending on the object and the method
(the best accuracy was achieved for half of the galaxies which have an active nuclei in the centre). 
 
The colliding star-forming Antennae galaxies constitute an obvious target for a detailed study 
of the separation between the X-ray sources and clusters, because they contain a couple of dozens 
bright X-ray sources \citep{Zezas2006} and a thousand catalogued  clusters \citep{WS95}. 
\citet{Zezas02b} have noticed that most of the brightest X-ray sources are displaced from the 
neighbouring star clusters. However, their absolute astrometry has an $\sim1\farcs5$ error ($\sim$ 160 pc), 
therefore they could not possibly measure displacements smaller than that  
and prove unambiguously an association between the  X-ray sources and clusters.  
Thus the question about the association of the X-ray sources and clusters remains open.

The aim of the present paper is to determine the distances between the brightest  
X-ray sources and the stellar clusters in the Antennae galaxies 
using an accurate astrometric solution (with error $<0\farcs3$). 
We find significant displacement between the  ULXs and the clusters as well as 
a highly significant association between them. We also estimate the ages of the stellar clusters. 
This allows us to determine the minimum masses of the ULX progenitors.

%%%%%%%%%%%%%%%%%%%%%%%%%%%%%%%%%%%%%
\section{Observations and data reduction}

\subsection{X-ray source sample}
 
The X-ray sources in the Antennae galaxies, NGC 4038/NGC 4039 (see Fig. \ref{fig:wholemap}),  
were surveyed by \citet{Zezas2002a, Zezas2006} and \citet{Swartz2004}.
From the catalogue of \citet{Zezas2002a} we have selected sources situated in the main bodies of the galaxies to 
reduce the chance  that they are background or foreground sources. 
We restrict the sample to those having more than 45 counts, which guaranties high accuracy of the coordinates. 
We then further down select the sources, which have average absorption corrected 
0.1--10~keV luminosities in excess of $2.75\times10^{38}$ erg~s$^{-1}$  as estimated by \citet{Zezas2006} 
assuming a power-law spectrum with photon index $\Gamma = 1.7$ and 
the Galactic line-of-sight hydrogen column density $N_{\rm H} = 3.4\times 10^{20} {\rm cm}^{-2}$ 
\citep{Stark92}.  
These luminosities were recomputed for the distance 
to the Antennae galaxies of 22~Mpc as recommended by \citet{Schweizer2008}. 
This distance gives  the scale of 107 pc/arcsec and the distance modulus $-31.71$. 

\begin{table}
\begin{center}
 \caption{The brightest X-ray sources in the Antennae galaxies and the nearest stellar clusters. } 
\label{tab:sources}
\begin{tabular}{cccccc}
\hline
Object$^a$  &  RA (J2000) &  Dec (J2000)     & $\log L_{\rm X}^b$ & Offset$^c$  & Cluster$^d$  
\\
\hline
\multicolumn{6}{c}{ULX sample} \\
%\hline
11  & 12:01:51.33 & $-$18:52:24.9   &  39.90  &  0.15  & 253    \\ 
						&   &     &   & 1.71 & 244  \\ 
						&   &     &   & 1.76 & 260        \\ 	
16  & 12:01:52.09 & $-$18:51:33.2   & 39.92 &  1.26    & 706  \\ 		
&   &     &   & 1.44 & 709        \\ 	
&   &     &   & 1.52 & 711   \\ 
&   &     &   & 1.56 & 707      \\ 	
&   &     &   & 1.68 & 708        \\ 	
29  & 12:01:53.50 & $-$18:53:10.5 & 39.57   &   0.30 & 49   \\ 
31  & 12:01:54.28 & $-$18:52:01.4   &  39.34  & 2.50  & 498  \\ 
						&   &     &   & 2.61  & 514    \\ 
 42  & 12:01:55.66 & $-$18:52:14.6   &  39.67 &  0.64  & 386   \\
&   &     &   & 0.67 & 374      \\ 	
&   &     &   & 0.90 & 388      \\ 
&   &     &   & 1.12 & 370      \\ 	
&   &     &   & 1.13 & 368      \\ 	
&   &     &   & 1.16 & 362         \\ 	
44  & 12:01:56.44 & $-$18 51 57.4   &  39.86  &  3.26  & 518  \\                                        
%\hline
\multicolumn{6}{c}{sub-ULX sample} \\
%\hline
5   & 12:01:50.48 & $-$18:52:15.4  &  38.57 &  1.42 & 387      \\ 	
&   &     &   & 1.48 & 381      \\ 	
&   &     &   & 1.54 & 377      \\ 		
6   & 12:01:50.52 & $-$18:52:04.2  &  38.59 &  0.36   & 491 \\
&   &     &   & 0.79 & 487      \\ 
&   &     &   & 1.08 & 483      \\ 	
&   &     &   & 1.13 & 495      \\ 	
&   &     &   & 1.36 & 485      \\ 	
&   &     &   & 1.44 & 479      \\ 	
&   &     &   & 1.67 & 481      \\ 	
13  & 12:01:51.63 & $-$18:52:31.3  &  39.09 &  2.02   & 206   \\ 
18  & 12:01:52.40 & $-$18:52:06.3   &  39.28  & 1.92   & --      \\
19  & 12:01:52.43 & $-$18:53:13.7   &  38.91 &  1.07  &  -- \\ 
20  & 12:01:52.75 & $-$18:51:29.5   & 38.48 &  0.30      & 720  \\
&   &     &   & 0.52 & 717      \\ 	
24  & 12:01:53.00 & $-$18:52:02.6   & 38.80  &  0.10      & 500  \\
&   &     &   & 0.37 & 501      \\ 	
&   &     &   & 0.72 & 503      \\ 	
25  & 12:01:53.01 & $-$18:52:09.1  & 38.44   &  0.67   & 450 \\
&   &     &   & 0.88 & 442     \\ 
&   &     &   & 1.40 & 451     \\ 	
32  & 12:01:54.36 & $-$18:52:09.8   & 38.96 &  1.44    & 443   \\ 
&   &     &   & 1.66 & 454      \\ 	
&   &     &   & 1.92 & 453      \\ 	
33  & 12:01:54.51 & $-$18:53:06.3   & 38.73  &  0.36    & 75  \\
&   &     &   & 0.67 & 77      \\ 	
&   &     &   & 1.25 & 65      \\ 	
34  & 12:01:54.56 & $-$18:53:02.7   & 38.90  &  0.63   & 90  \\
&   &     &   & 0.93 & 89      \\ 
&   &     &   & 1.23 & 87    \\ 
&   &     &   & 1.59 & 88      \\ 	
&   &     &   & 1.84 & 86      \\ 	
35  & 12:01:54.78 & $-$18:52:51.9   &  39.09  &  0.57   & 115   \\ 
&  &   &    &  0.83   & --    \\
&   &     &   & 1.95 & 109      \\ 	
41  & 12:01:55.49 & $-$18:52:35.4  & 38.68  &  2.24   & 161 \\
46  & 12:01:56.65 & $-$18:52:00.4  & 38.45  &  2.33     & 507  \\
\hline
\end{tabular}
%\end{minipage}
\end{center}
\begin{flushleft}{
$^{a}${Source number according to \citet{Zezas2002a}.} 
$^{b}${Logarithm of luminosity in erg s$^{-1}$ in the range 0.1--10~keV \citep{Zezas2006} 
corrected for absorption and recomputed for the distance of 22 Mpc. }  
$^{c}${Offset (in arcsec; here 1\arcsec=107 pc) between the X-ray source and the  stellar cluster. }  
$^{d}${Cluster ID from the WS95 catalogue if available.    
}
}\end{flushleft} 
\end{table}

We further remove from the sample source X-37, which is a background quasar \citep{Clark05} 
and will be further used for accurate astrometry. This leaves us with 20 sources (see Table \ref{tab:sources}). 
Among these sources, X-24 and X-29 are situated very close to the northern and southern nuclei, 
with the offsets from the bright 2MASS nuclei of 1\farcs6 and 0\farcs3, respectively. 
Thus, with the worse astrometry, we could have assumed them to be the black holes 
in the galactic centers. However, this is not the case. 
There are no indications of any AGN activity in the nuclei \citep{Zezas02b,Ueda2012}. 
Six  sources with luminosities $L_{\rm X} \gtrsim 2\times 10^{39}$~erg~s$^{-1}$   \citep{Zezas2002a,Zezas2006,Swartz2004} 
may be considered as bona fide ULXs (the ``ULX sample''). 
We call other sources with luminosities in the range 
$3\times10^{38} \lesssim L_{\rm X} \lesssim 2\times 10^{39}$~erg~s$^{-1}$ 
the ``sub-ULX sample''. 
All the sources are located in the main bodies of the galaxies, however, a few sources (X-18, X-19) 
are located in the very obscured dusty regions.

\begin{figure*}
\includegraphics[width=16.0cm]{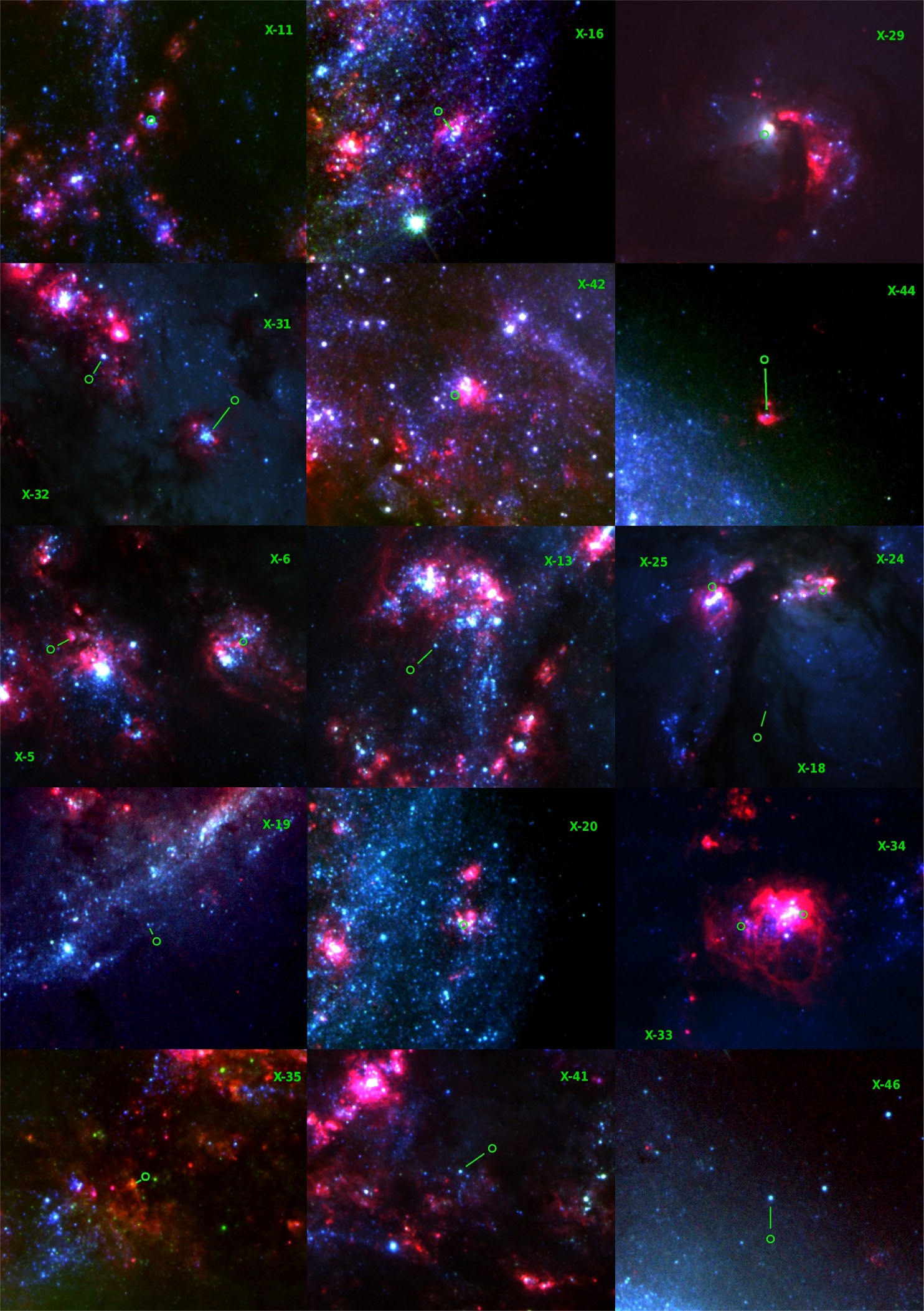}
\caption{Enlarged $18\arcsec\times 15\arcsec$ (1\arcsec=107 pc)  rgb images 
from Fig. \ref{fig:wholemap} for the regions around the brightest X-ray sources. 
The images were produced from ACS images taken in FR656N (H$\alpha$, red), F435W (blue) and F814W (green) bands. 
Chandra positions (0\farcs2 radius) of the X-ray sources are indicated in green. 
The bar indicates the cluster (with $V<24$) closest to an X-ray source, when it is not obvious.  }
\label{fig:ulx6}
\end{figure*}
%%%%

\subsection{Optical observations}
\label{sec:optical}

For spectral study we have selected regions around four ULX  X-11, 16, 31, 42 and also X-35. 
As X-32 is situated very close to X-31, it was also studied. 
We did not observe X-44, because 
it lies at the outskirts of the galaxies and we did not originally expect bright optical sources around it. 
Source X-29 was not observed, because it seemed to be associated with the nucleus of NGC 4039 
(we, however, keep it in the ULX sample). 
We obtained two data sets on 2007 February  and 2008 March  with the Visible Multi-Object Spectrograph (VIMOS) 
Integral Field Unit (IFU) at the 8.2m Very Large Telescope Melipal at ESO's Paranal Observatory in Chile. 

The VIMOS IFU consists of 1600 (40$\times$40) fibres coupled to microlenses and operates with four $2048\times4096$ 
EEV44 CCD detectors yielding four quadrants \citep{Vimos2005}.  
The field-of-view of 13\farcs{}5$\times$13\farcs5 was covered by 1600 spatial pixels (spaxels) at a spatial sampling of 
0\farcs{}33 per spaxel. Two different grisms were used, high resolution blue  ($\sim$0.51\,\AA/pixel), 
covering the wavelength range of 4150--6200\,\AA\AA\ and high resolution orange ($\sim$0.60\,\AA/pixel) 
covering the range  of 5250--7400\,\AA\AA. 
The spectral resolution measured using the sky emission lines is 1.86  and 1.88\,\AA\ for the blue and orange grisms, respectively. 
We have taken  spectral images of each target with exposure times of 6$\times$900~s in the blue and 
4$\times$900~s  in the red regions.

For each VIMOS quadrant all the spectra were traced, identified, bias subtracted, flat field corrected, 
corrected for relative  fiber transmission, and wavelength calibrated using the routines of the ESO Recipe 
EXecution pipeline ({\tt ESOrex}).\footnote{{\tt ESOrex} is developed and maintained by the European Southern Observatory.} 
We have checked the fiber traces in the images and corrected them, when it was necessary, using bright night 
sky and nebular lines. Cosmic rays and bad pixels were identified and cleaned using standard {\tt IRAF}
\footnote{{\tt IRAF} is distributed by the National Optical Astronomy Observatory, which is 
operated by the Association of Universities for Research in Astronomy (AURA) under cooperative agreement 
with the National Science Foundation.} routines. 
Finally, the intensity of the night-sky emission line [O\,{\sc i}]$\lambda 5577$ was used to correct for the different relative transmission of the VIMOS quadrants.

The processed spectra were organized in data cubes using the tabulated correspondence between 
each fiber and its position in the field of view. 
The spectra were co-added after correcting for the position offset. 
The accuracy of the offset is $\pm$0.2 pixel ($\simeq$0\farcs07). 
This slight deterioration of the spatial resolution does not affect the results.

We use also archival imaging observations taken with the {\it Hubble Space Telescope} ({\it HST}). 
The ACS-WFC images taken in F435W, F550M and F814W filters on  2005 July 21 and images 
in FR656N filter taken on 2005 June 29  were used for astrometry 
and for illustrations (Figs \ref{fig:wholemap} and \ref{fig:ulx6}). 
The WFPC2 images taken on 1996 January 20 in F336W, F439W, F555W and F814W filters \citep{Whitmore99} 
were used for photometry of the clusters associated with the ULXs. 
The WFPC2 images were reduced in a standard way with {\tt hstphot} \citep{HSTphot2000}. 
The same routine was used to measure all point-like sources in the images 
and to calibrate them in the standard $U$, $B$, $V$ and $I$ magnitudes \citep{Johnson66}. 
Using standard {\tt IRAF} tools we carried out aperture photometry of stellar clusters presumably 
associated with the ULXs. 
Corrections of the WFPC2 magnitudes for interstellar extinction were made according to \citet{Holtzman95}.

%%%%
\begin{figure}
\includegraphics[width=8cm]{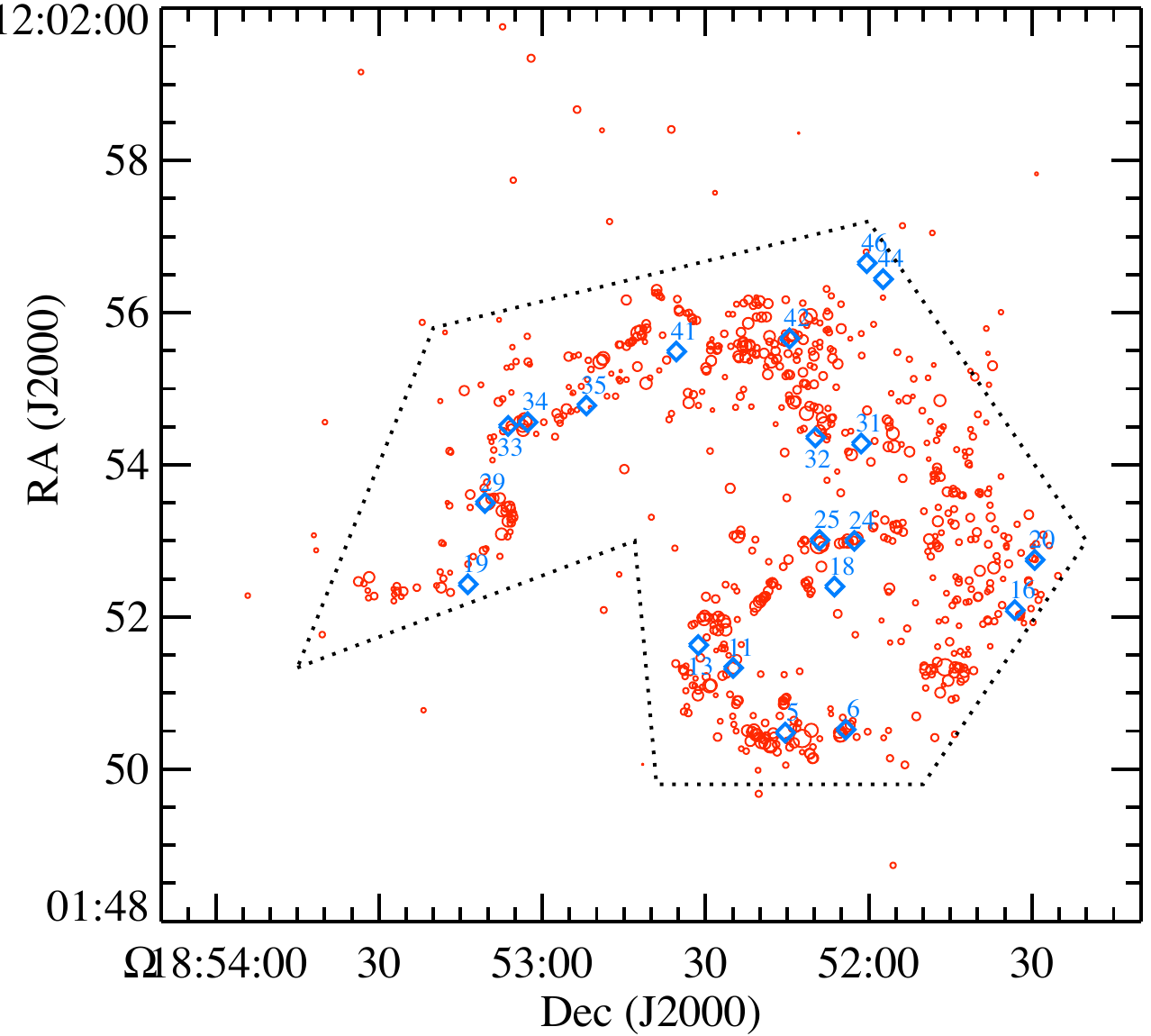}
\caption{Positions of the bright X-ray sources (diamonds) 
and the stellar clusters from the WS95 catalogue (red circles).  
The X-ray sources are numbered from the catalogue of \citet{Zezas2002a}. 
The size of the circle indicates the brightness of the cluster. 
The dots outline the main body of the galaxies. 
} 
\label{fig:clusters_xray}
\end{figure}
 %%%%

\subsection{Astrometry of X-ray sources and stellar clusters}

To have an accurate absolute astrometry in optical images we used the 
archival r-band images   taken with the MegaCam at the Canada-France-Hawaii Telescope (CFHT).  
For an astrometric solution, we used 15 point-like 2MASS \citep{2MASS2003} sources surrounding the Antennae galaxies in $4\arcmin\times5 \arcmin$ field of the CFHT images. The accuracy of coordinates (rms) is better than 0\farcs15. 
In the next step we transfered this solution from CFHT to ACS images of the galaxies with 22 common point-like objects, 
the accuracy of this transfer is much better than 0\farcs15. 
The same has been done with the WFPC2 images of the galaxies. We used  a routine described by \citet{Gooch97}. 
Our final accuracy in absolute astrometry of the optical {\it HST} images is the same as that in 2MASS, i.e. about 0\farcs15.

The brightest 12 X-ray sources have $>$100 net source counts 
and the rest of the sources still have $>$45~cts in individual {\it Chandra} images. 
They are point-like sources and the internal accuracy of their coordinates  
$\approx 1\arcsec/\sqrt{\mbox{counts}}$ is expected to be below 0\farcs15 \citep{Swartz2004}. 
For the ULX-sample, we have averaged the source coordinates presented in surveys by \citet{Zezas2002a} and \citet{Swartz2004}. 
The ULX coordinate difference in these two surveys is 
below 0\farcs2  with one exception for X-35 where it is $\approx$0\farcs3.
For the sub-ULX-sample we adopt the coordinates from \citet{Zezas2002a}.
The final internal accuracy of the  X-ray coordinates is better than $\approx$0\farcs2.  
To determine the offset between the X-ray and optical images we used a background quasar \citep{Clark05}, 
the source X-37 in the Antennae galaxies, which is a very bright point-like source both in the X-ray and optical images. 
The coordinates of the X-ray sources  corrected for this offset are presented in Table~\ref{tab:sources}. 
The final coordinate accuracy for the {\it HST}/ACS -- {\it Chandra}  solution 
is $\approx$0\farcs2--0\farcs25, depending on the brightness of the source.

\citet[][ hereafter WS95]{WS95} present a catalogue of stellar clusters obtained using {\it HST}/WFPC1  
with the coordinates relative to the source N442. 
We recomputed the  coordinates of all the clusters using our solution. 
For all clusters in the vicinity of X-ray sources (see Table~\ref{tab:sources}) 
we have obtained independent measurements of the coordinates using 
ACS images and found the rms displacement from the corrected WS95 values of 0\farcs12. 
This allows us to use the whole WS95 catalogue with the corrected coordinates for  the statistical 
studies and modelling of the distribution of the offsets between X-ray sources and clusters.

%%%%
\begin{figure}
\begin{center}
\includegraphics[width=7cm]{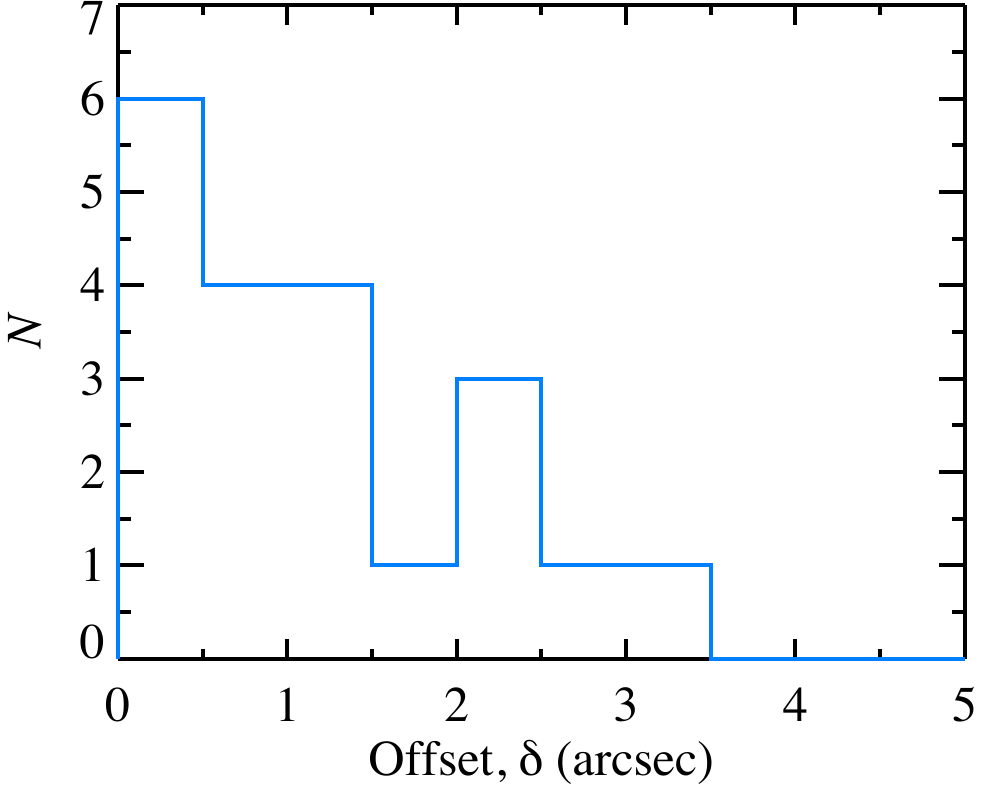}
\end{center}
\caption{Distribution of the offsets (1\arcsec=107 pc) between the positions of 
the X-ray sources and the optical positions of nearest stellar clusters (Table~\ref{tab:sources}). 
}  
\label{fig:histo}
\end{figure}
 %%%%

%%%%%%%%%%%%%%%%%%%%%%%%%%%%%%%%%%%%%%
\section{Association  with clusters}   

In Fig.~\ref{fig:wholemap} we show rgb image of the Antennae galaxies with the brightest X-ray sources.
In Fig.~\ref{fig:ulx6} we show enlarged rgb images around the X-ray  sources. 
The positions of stellar clusters  from WS95 catalogue together with the X-ray sources from Table \ref{tab:sources} 
are shown in Fig. \ref{fig:clusters_xray}. 
A remarkable property of the X-ray sources is they are located close to young stellar clusters and complexes.
Only a few sources (X-11, X-20, X-24, X-29) can be claimed to coincide with the position of stellar clusters 
(within 0\farcs{}3), for  other sources the distances to the clusters is above the astrometric accuracy. 
In Fig.~\ref{fig:histo} we show distributions of the offsets between the positions of the X-ray sources and the nearest stellar clusters.
All the sources have separation below 2\farcs7 ($\sim$290~pc), except X-44 ($\sim$350~pc).
Source  X-18 is located in a heavily obscured region and X-19 is at the edge of the star formation belt, 
with the nearest bright WS95 clusters more that 4\arcsec way. 
However, for both these sources there exist much closer non-catalogued clusters (see Fig.~\ref{fig:ulx6}).

The displacement between the X-ray source positions and optical clusters in Antennae was noticed 
by \citet{Zezas02b} and discussed by \citet{ZezFab02}. 
However, they have used the absolute astrometry of the WFPC2 images, which has an $\sim1\farcs5$ error, 
and considered X-ray sources to coincide with the optical sources if their separation is less than 2$\arcsec$. 
Our accurate astrometry does not leave any doubts that the offsets of 0\farcs5--3\arcsec are real.

\begin{table}
\begin{center}
\caption{Correlation between stellar clusters and X-ray sources. }
\label{tab:correlation}
\begin{tabular}{ccccccc}
\hline
$V_{\max}$$^a$  & $N_{\rm cl}$$^b$  & $\delta$ (\arcsec)$^c$ & $n_{\rm e}$$^d$ &  $n$$^e$ 
& $p$ ($10^{-7}$)$^f$ & $N_{\rm X,cl}$$^g$ \\
\hline
21.5  &  259 & 1.0 & 1.8   & 15 & 0.09 & 8 \\
         &         & 1.5 & 4.1   & 26 & 0.31 & 10 \\
         &         & 2.0 & 7.2   & 35 & 0.63 & 11 \\
%         &         & 2.5 & 11.8 & 40 & 1.4 \\
22.0  &  371 & 1.0 & 2.6   & 17 & 0.18 & 9 \\ 
         &         & 1.5 & 5.8   & 32 & 0.5 & 11 \\
         &         & 2.0 & 10.3 & 42 & 1.1 & 12 \\
%         &         & 2.5 & 16.7 & 49 & 2.4 \\		               
22.5  &  535 & 1.0 & 3.7   & 18 & 1.2 & 9 \\ 
         &         & 1.5 & 8.4   & 35 & 0.8 & 11 \\
         &         & 2.0 & 14.9 & 48 & 2.0 & 12 \\
%         &         & 2.5 & 24.4 & 60 & 4.0\\
24.5  &  690 & 1.0 & 4.8   & 20 & 2.3 & 10\\  	
         &         & 1.5 & 10.8 & 37 & 1.2 & 12 \\
         &         & 2.0 & 19.2 & 52 & 2.9 & 13 \\
%         &         & 2.5 & 31.5 & 67 & 5.3 \\	
\hline
\end{tabular}
\end{center}
\begin{flushleft}{
$^{a}${Cut in $V$ magnitude of the WS95 catalogue.} 
$^{b}${Number of clusters from the WS95 catalogue brighter than $V_{\max}$. }
$^{c}${Maximum offset.} 
$^{d}${Expectation number given by equation (\ref{eq:exp_value}).} 
$^{e}${Number of clusters within distance $\delta$ from all X-ray sources.} 
$^{f}${Poisson probability that the association between clusters and X-ray sources is random.} 
$^{g}${Number of X-ray sources, which have at least one cluster within  $\delta$.}  
}
\end{flushleft} 
\end{table}

%%%%
\begin{figure}
\begin{center}
\includegraphics[width=7cm]{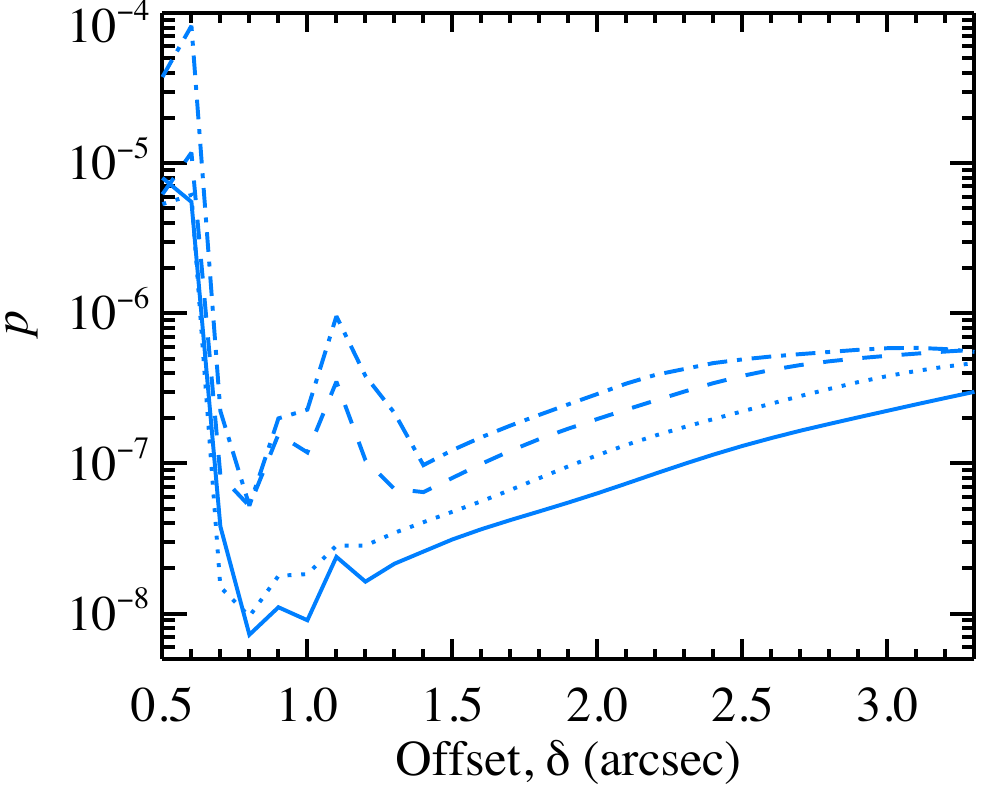}
\end{center}
\caption{Probability that the association between clusters and X-ray sources is random 
as a function of the offset angle (1\arcsec=107 pc). 
The Poisson probabilities computed from the two-point correlation function for various 
WS95 sub-catalogues (clusters lying within the area identified in Fig.~\ref{fig:clusters_xray})
with the cuts in visual magnitude  $V_{\max}$=21.5, 22.0, 22.5 and 24.5
are  shown by solid, dotted, dashed and dot-dashed lines, respectively.   
}
\label{fig:n_corr}
\end{figure}
 %%%%

To estimate the statistical significance of the clustering of the bright X-ray sources near the stellar clusters we 
first use the usual two-point correlation function, counting the number $n$ 
of clusters from a given subset of the WS95 catalogue within an angle $\delta$ from each X-ray source. 
We then compare this number to the expectation $n_{\rm e}$ for the null hypothesis that the X-ray sources are 
distributed randomly over the galaxies bodies  
\begin{equation} \label{eq:exp_value}
n_{\rm e} =  N_{\rm cl}(V_{\max}) N_{\rm X}\ \pi \delta^2 /\Sigma , 
\end{equation}
where $N_{\rm cl}(V_{\max}) $ is the number of clusters of magnitude below the cut $V<V_{\max}$,
$N_{\rm X}$ is the number of X-ray sources, and $\Sigma\sim 9000 \Box\arcsec$
is the area  containing stellar clusters (see Fig.~\ref{fig:clusters_xray}).
We introduce the measure of the signal $p$ as the probability of sampling $n$ or more hits from the Poisson distribution at expectation $n_{\rm e}$. This probability depends on $\delta$ and cuts in the stellar cluster catalogue $V_{\max}$.  
The null hypothesis is rejected with a very high significance $p<10^{-4}$ for any tried $\delta$ and 
any reasonable cut $V_{\max}$ (see Fig.~\ref{fig:n_corr} and Table~\ref{tab:correlation}). 
It reaches the minimum at the $10^{-8}$ level for the correlation radius  $\delta\sim$0\farcs7--1\farcs5.  
The deepest minimum corresponding to the cuts $V_{\max}=21.5$ and 22.0
indicates that the strongest correlation exists between X-ray sources and the brightest clusters.  
About half of the X-ray sources have at least one cluster within 1\arcsec--2\arcsec (see Table~\ref{tab:correlation}).
Interestingly, that 15 out of 18 X-ray sources and all ULXs 
have nearest WS95 clusters brighter than $V = 22.3$. 
Besides that, practically all the clusters and cluster complexes associated with the X-ray sources 
(i.e. the closest WS95 clusters) are young (see Fig.~\ref{fig:ulx6}), less than 10~Myrs, 
because they are surrounded by strong H$\alpha$ emission \citep{Starburst99}. 
Because the fraction of so young clusters in the WS95 catalogue 
is below 30 per cent \citep{Fall05}, the correlation between the X-ray sources and 
the young clusters  is even stronger.

We can also check the hypothesis that the X-ray sources are distributed according to the 
stellar density \citep{Rangelov11}. 
The stars, however, are distributed rather homogeneously over a larger area than the stellar clusters \citep{Whitmore99}
and therefore the significance of the association would be even larger.

\begin{figure}
\begin{center}
\includegraphics[width=7cm]{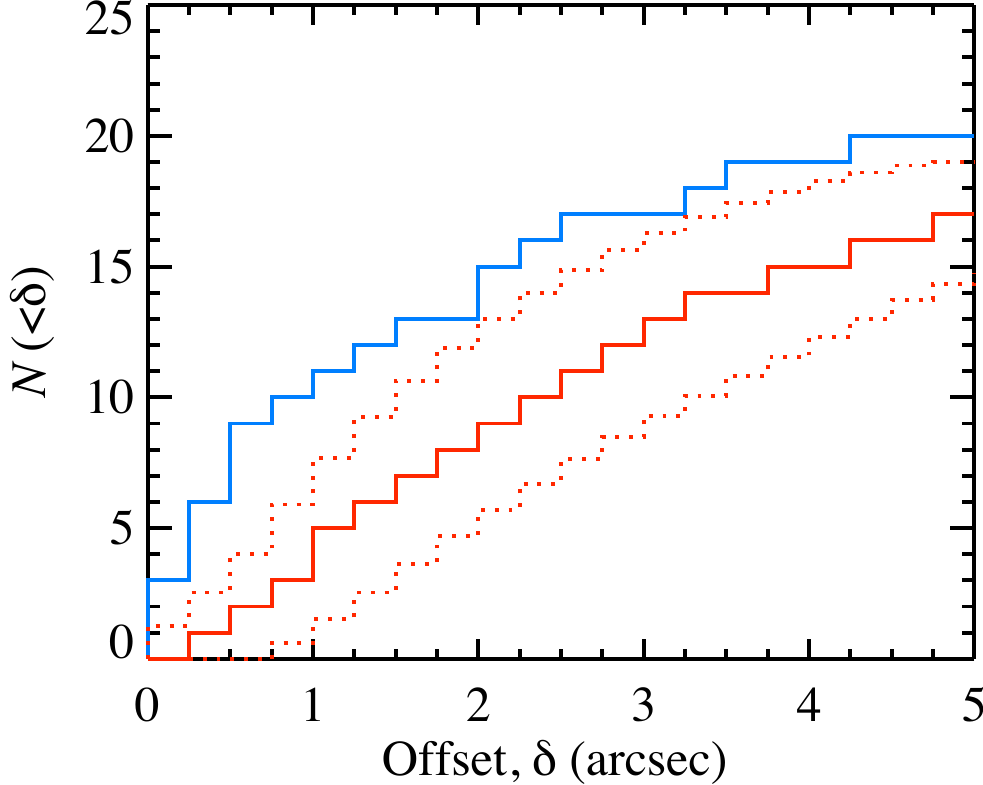}
\end{center}
\caption{Cumulative distribution of the displacements between X-ray sources  and 
stellar clusters from the WS95 catalogue (upper  blue solid histogram). 
The lower red solid histogram shows the median distribution obtained from  $10^5$ 
catalogues of X-ray sources distributed randomly over the galaxies body within 
the limits shown in Fig. \ref{fig:clusters_xray}. 
The dotted lines show the limits that enclose 90 per cent of the simulated distributions 
in each bin separately.  
}
\label{fig:cumul}
\end{figure}

Fig.~\ref{fig:cumul} shows the cumulative distributions of the offsets between the  X-ray sources and stellar clusters
from WS95 catalogue and those obtained from artificial X-ray source samples scattered 
randomly over the main body of the galaxies (limited by the dotted line in Fig.~\ref{fig:clusters_xray}).
We see that the observed distribution is significantly higher 
than the median of simulated offset distributions 
and  even the  distribution corresponding to the 90\% confidence limit. 
Specifically, the observed distribution has about twice as much sources within 1\farcs5 than the simulated one. 
This confirms the association between bright X-ray sources and stellar clusters.

%%%%
 %%%%

\begin{figure}
\includegraphics[width=8cm]{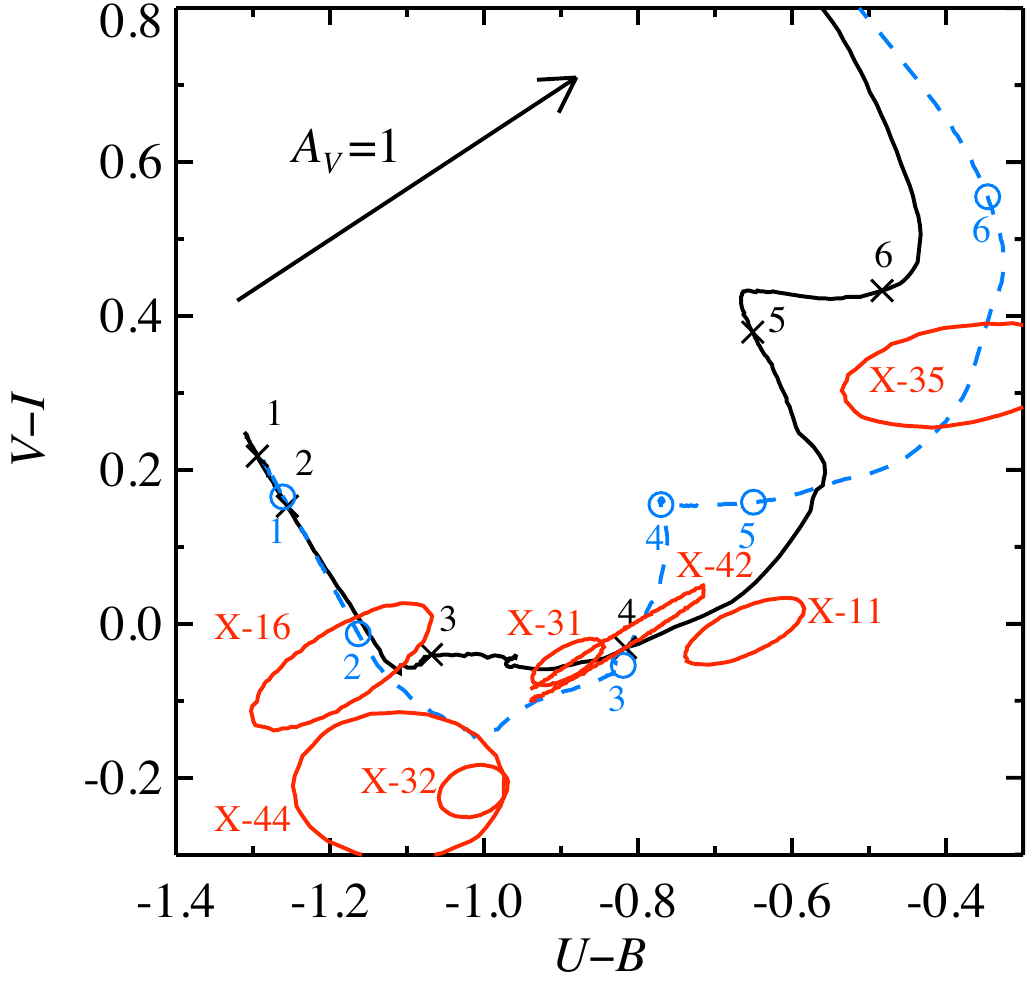}
\caption{Colour-colour  
diagram  for the studied clusters together with the Starburst99 \citep{Starburst99} 
evolutionary tracks from the Geneva group. 
The tracks are for instantaneous starburst model with the \citet{KroupaIMF01} IMF 
extending to $150\msun$ 
for solar metallicity $Z=0.02$ (solid line) and for double-solar metallicity (dashed line). 
Black crosses and blue circles on the tracks and the numbers next to them 
indicate the age in million years (from 1 to 6~Myrs). 
The arrow shows a direction of the shift due to the interstellar extinction. 
The extinction corrected positions of the studied clusters  are shown by ellipses. 
They correspond to 68 per cent confidence limit computed using Monte-Carlo simulations 
from the combined uncertainties in the colours and  $A_V$.  
}
\label{fig:hrdiag_starburst}
\end{figure}

\begin{table*}
\begin{center}
 \begin{minipage}{150mm}
 \caption{Properties of the stellar clusters next to the  brightest X-ray sources in the Antennae galaxies and 
 the minimum masses of ULX progenitors. }
\label{tab:clusters}
\begin{tabular}{ccccccccccc}
\hline
Object$^a$  & Cluster$^b$  & $Z/Z_{\sun}$$^c$ & $A_V$$^d$ & $A_{V}^*$$^e$ & $V$$^f$ & $M_V$$^g$  & Age$^h$  & $M_{\rm clus}$$^i$ & $v_{\rm ej}$$^j$  & $M_{\rm prog}$$^k$ \\
                      &                 & &   &                   &                &                       
                      & Myr  &   $10^5{\msun}$   & km~s$^{-1}$ & $\msun$\\
\hline     
11  & 253  & 1.0 & 0.98$\pm$0.09 & 0.98 & 18.80  & $-$12.91   & 4.2$\pm$0.1 & 0.99  & -- & 50 \\   
16  & 706  & 1.0 &  0.60$\pm$0.16  & 0.52 & 19.71  &  $-$12.00  & 2.6$\pm$0.2 & 0.59  & 50 &110 \\ 	                
16  & 706  & 2.0 &  0.60$\pm$0.16  & 0.52 & 19.71  &  $-$12.00  & 2.1$\pm$0.2 & 0.62  & 61 &100 \\ 	
31  & 498  & 1.0 & 0.40$\pm$0.06 & 0.46  & 18.51  &  $-$13.20   & 3.8$\pm$0.1 & 1.67  & 67 &60  \\                      
31  & 498  & 2.0 & 0.40$\pm$0.06 & 0.46  & 18.51  &  $-$13.20   & 2.8$\pm$0.2 & 0.91  & 91 & 100  \\    
32  & 443  & 2.0 & 0.94$\pm$0.04 & 0.94  & 18.59  & $-$13.12   & 2.5$\pm$0.1 & 1.12  & 59 & 100 \\     
35  &  --   & 2.0 & 3.43$\pm$0.11 & 3.40  & 18.80  & $-$12.91  & 5.7$\pm$0.2 & 1.00  & 15 & 30 \\  
42  & 386  & 1.0 & 0.46$\pm$0.27 & 0.58  & 17.14 & $-$14.57   & 3.9$\pm$0.1 & 5.66   & 17 & 60  \\
42  & 386  & 2.0 & 0.46$\pm$0.27 & 0.40  & 17.32  & $-$14.39  & 3.3$\pm$0.1 & 3.82   & 20 & 70  \\
44  & 518  & 2.0  &  --   & 0.155   &  21.96   & $-$9.75    & 2.4$\pm$0.1 & 0.055  & 139 & 100  \\ 
\hline  
\end{tabular}
\end{minipage}
\end{center}
\begin{flushleft}{
$^{a}${Source number according to \citet{Zezas2002a}.} 
$^{b}${Nearest stellar clusters and their ID from the WS95 catalogue if available.  }
$^{c}${Metallicity in solar units.} 
$^{d}${Mean extinction value and standard deviation determined from the VIMOS spectra 
in the field of $3\times3$ pixels at the position of the cluster.  } 
$^{e}${Extinction best matching the evolutionary tracks.} 
$^{f}${Dereddened WFPC2 $V$ magnitude.} 
$^{g}${Absolute $V$ magnitude for distance of 22 Mpc.}  
$^{h}${Age of the stellar cluster estimated from Fig.~\ref{fig:hrdiag_starburst}. }
$^{i}${Mass of the stellar cluster computed from 
the difference between the observed $M_V$ given in this Table  
and the theoretical value from \citet{Starburst99} 
at the best-fitting age and metallicity determined from Fig.~\ref{fig:hrdiag_starburst}.}  
$^{j}${The minimum ejection velocity (not accounting for deprojection) of the X-ray source
needed to reach the required displacement from the cluster at a given age. }
$^{k}${Minimum mass of the ULX progenitor in ${\msun}$ obtained from maximum possible age of the cluster
using the Geneva group stellar evolution models as described in Starburst99. }
}
\end{flushleft} 
\end{table*}

\section{Cluster ages and the masses of ULX progenitors}

The VIMOS images (Fig. \ref{fig:wholemap}) cover practically all the ULX sources in the Antenna galaxies. 
We first study the spectra of the nebulae emission in the VIMOS fields around 
the clusters (see Table~\ref{tab:sources} and Figs~\ref{fig:6mapsX11}--\ref{fig:6mapsX42}). 
We determine the H$\alpha$/H$\beta$ flux ratios and compare them to the theoretical value of 2.87 
corresponding to the case B of  gaseous nebulae (which is the same to within 10 per cent in a wide range of 
temperatures and densities, see \citealt{Osterbrock_Ferl2006}).
This allows us to find the extinction values all over the fields 
(see Table~\ref{tab:sources} and the extinction panels in Figs~\ref{fig:6mapsX11}--\ref{fig:6mapsX42}). 
The  $A_V$ measured close the X-ray sources are in perfect agreement with the independent estimates 
from the hydrogen column density $A_V=5.5\  N_{\rm H}/10^{22}\ {\rm cm}^2$ \citep{PS95}
obtained from the {\it Chandra} spectra \citep[see Table 5 in ][]{Zezas2002a}.  
For cluster N518 next to X-44 we do not have VIMOS images and we 
use the Galactic extinction $A_V=0.155$, which is the minimum possible value and 
is close to the maximum value consistent with the evolutionary tracks (see below).

The dereddened $U$, $B$, $V$ and $I$ magnitudes for the clusters were obtained 
from the  WFPC2 images  as described in Sect.~\ref{sec:optical}. 
On Fig.~\ref{fig:hrdiag_starburst}, we show the colour-colour $V-I$ and $U-B$  diagram with 
the error contours for each cluster, which are elongated along the vector $A_V$.  
These colours are then compared  to the cluster evolution tracks from the Geneva group
computed using Starburst99 code\footnote{Starburst99 is available at http://www.stsci.edu/science/starburst99/} 
\citep{Starburst99,VL05}. 
We consider the instantaneous starburst model with the \citet{KroupaIMF01} initial mass function (IMF) 
extending to $150\msun$
for solar and double-solar metallicities (see Fig.~\ref{fig:hrdiag_starburst}).
The cluster next to X-11 is consistent with the solar metallicity and those next to X-32, 35 and X-44 
with only  double-solar metallicity. 
On the other hand, the colours of the clusters next to X-16, 31 and 42 can be described by any metallicity 
from solar to double-solar, which  is consistent with the range of metallicities from $0.9Z_{\sun}$ to $1.3Z_{\sun}$
found by  \citet{Bastian09}  in their study of 16 young clusters in the Antennae.
For the cluster next to X-44 we took the Galactic extinction and 
increasing the value of $A_V$ would shift the cluster even further away from the
theoretical tracks to the lower-left corner of the colour-colour diagram. 

The overlap at the colour-colour diagram between the cluster error contours  
and the evolutionary tracks gives us an  estimate of the cluster ages. 
We find that all clusters associated with the ULXs are extremely young with the age of less than 6~Myr  
(see Table~\ref{tab:clusters}).\footnote{For the cluster next to X-35 there is 
another solution with the age of $\sim$20 Myr, which we reject, because of the presence of the 
strong H$\alpha$ emission. } 
For the assumed metallicity,  an improved estimate for $A_V$ can be obtained from the same overlap.  
Comparing thus obtained $M_V$ to the theoretical absolute magnitude from the Starburst99 models 
(for a cluster of mass $10^6\msun$), we find  the cluster masses using equation 
$\Delta M_V=-2.5\log (M_{\rm clus}/10^6\msun)$ (see Table~\ref{tab:clusters}), 
which vary between $5\times10^3$ and $6\times10^5\msun$. 
For the clusters which are consistent with both metallicities, the difference in the ages 
gives us an estimate of the systematic error.

In the X-31/X-32 VIMOS image (see Fig. \ref{fig:6mapsX32}) there are also two bright stellar complexes 
N455 and N418 previously studied by \citet{Bastian06} (named 5 and 6 there) and showing spectra with strong 
Wolf-Rayet emission features. This confirms the young age ($\lesssim 4$~Myr) of the clusters in that region. 
In Fig.~\ref{fig:spectra}  we present VIMOS spectra of the clusters studied (together with spectra of the 
complexes N\,5 and N\,6) demonstrating very strong hydrogen, [O\,{\sc iii}], [Ar\,{\sc iii}] and other nebular 
lines, which demand strong photoionizing continua of the clusters and confirm their young age. 
However, the cluster age estimates from the photometric data are notably more accurate than those 
obtained from the spectroscopy.  
We note that the independent evaluation of $A_V$ from the spectra  
break the known $A_V$--age degeneracy  \citep[see e.g.][]{Bastian09}.

The clear association of the X-ray sources both in the ULX and sub-ULX samples  with 
the young star clusters in the Antennae galaxies indicates that these sources  originate from 
the massive binaries ejected from the star clusters. 
The extremely young ages of the clusters associated with the ULXs put the lower limit on the mass of 
their progenitor stars, 
which varies between 30 and 110 $\msun$ according to the Geneva stellar evolution 
models used in \citet{Starburst99} (see Table \ref{tab:clusters}), 
indicating that all studied ULXs in the Antennae galaxies are associated with the most massive stars.  
 
The distribution of displacements and the cluster age allows us to give 
an estimate for the minimum ejection velocity of the sources from the clusters. 
Among the ULX sample (Table~\ref{tab:clusters}) only one source X-11 resides in a cluster. 
Other six sources have an average ejection velocity (accounting for deprojection, which gives 
a factor of $4/\pi$) of 77~km~s$^{-1}$.  
As X-44 is situated rather far away from the suspected parent cluster and it requires 
a very large ejection velocity (for a given small cluster mass), their association is probably spurious.

\section{Discussion}  

The correlation between  X-ray source positions and stellar clusters was already noticed by \citet{Kaaret04} 
for three starburst galaxies M\,82, NGC\,1569 and NGC\,5253 (which lie closer than 4~Mpc).  
In spite of their astrometric accuracy of $\ga$1\arcsec,  the proximity of the galaxies permitted them also 
to find the displacement between X-ray sources (with X-ray luminosities less than $10^{39}$~erg s$^{-1}$) and bright clusters. 
They found that brighter X-ray sources preferentially occur closer to the clusters, and no 
sources at luminosities above $10^{38}$~erg s$^{-1}$ are displaced by more than 200~pc. 
The lower luminosity sources are mostly low-mass X-ray binaries, which totally lose 
connection to their parent clusters because of their large ages
or because the clusters disappear on that time frame.  
In their study only one galaxy (M\,82) has sources brighter than $10^{38}$~erg s$^{-1}$ and only one source was brighter than $10^{39}$~erg s$^{-1}$. 

Recently \citet{Rangelov11} found that bright sources (with luminosities 
$3\times10^{36}$--$8\times10^{38}$~erg s$^{-1}$)  
in NGC 4449 are located close to the young stellar clusters with the age less than 6--8 Myr. 
This association was, however, only marginal, when they assumed that X-ray sources follow the light of the galaxy.
It is worth mentioning that about half of the X-ray sources  in NGC 4449 are probably low-mass X-ray binaries 
and the correlation between the remaining candidate high-mass X-ray binaries and stellar clusters is significant.  

The  Antennae galaxies have a relatively high star-formation rate and 
most of bright X-ray sources are probably high-mass X-ray binaries. 
\citet{Clark11} identified there 38 infrared counterparts within the 1\farcs0 
error boxes of 120 X-ray sources from the \citet{Zezas2006} catalogue. 
The majority of these IR sources are star clusters. 
For ten of them they were able to measure masses of about $10^6\msun$ and 
ages, which lie in a very narrow range around $10^7$ yr.
Their study, however, was limited to the immediate surrounding of the X-ray sources
and therefore it missed many clusters that still could be the birthplaces of the X-ray sources.  
The recent study by \citet{Rangelov12} showed that 22 out of 82 X-ray sources from the same catalogue 
have star clusters as optical counterparts within 1\farcs1 error box.\footnote{Their astrometry has 
1$\sigma$ error of 0\farcs54, but we note that 
the difference between the  {\it HST} and {\it Chandra} coordinates 
of the QSO X-37 is certainly much smaller than quoted value of the mean shift of 2\farcs5175 
obtained by \citet{Rangelov12}. }
They have measured the age of these clusters and the extinction using $UBVI$ and H$\alpha$ magnitudes 
finding that 14 sources are situated next to young clusters with ages of 4--6 Myr. 
Among the clusters associated with ULXs, they found the age of 5 Myr
for X-11 (their source X27), which is in agreement with our estimate of $4.2\pm0.1$ Myr. 
However, for  X-42 (their X99), they find the age of 200 Myr, while we find it to be below 4 Myr. 
This discrepancy probably results from the $A_V$--age degeneracy in colour-colour diagrams 
\citep[see e.g.][]{Bastian09}. 
On the other hand, our measurements of the $A_V$ are independent of the photometry. 
Similarly to \citet{Clark11}, \citet{Rangelov12} have 
limited themselves to the clusters very close to the X-ray sources, which 
have led them to an erroneous conclusion  that black holes stay within the clusters.

Neither \citet{Clark11}  nor  \citet{Rangelov12}  have considered a possibility that the clusters 
and the X-ray sources separated by more that 1\arcsec\ are related. 
We, however, find that majority of the X-ray sources  in the Antennae galaxies with 
luminosities $\gtrsim 3 \times 10^{38}$~erg s$^{-1}$ are located close to (but not within) very young clusters 
and there is a significant association between these types of sources. 
Together with findings by \citet{Kaaret04} this means that the association 
with clusters and the displacements are the common properties of all high-mass X-ray binaries and the ULXs 
do not differ in this aspect from the less bright X-ray binaries.
The correlation and the displacement of the brightest X-ray sources up to 300 pc from the clusters imply 
that they have been ejected from the clusters. It is not possible that the X-ray sources 
have been left behind their dissolved host clusters, because the extremely small ages of the clusters 
do not leave any time for the cluster (and nearby clusters) disintegration. 

There exist three known runaway scenarios, which can  explain these displacements: 
\begin{enumerate}
\item Instantaneous, symmetric mass loss in a supernova explosion in a close binary 
\citep{Zwicky57,Blaauw61,Shklovskii76,Tauris98,vandenHeuvel2000};
\item Instantaneous momentum impulse, or ``kick'', imparted through asymmetry of the supernova 
explosion in formation of a neutron star \citep{Woosley87,CordesChernoff98}. 
A non-radial asymmetry in neutrino momentum distribution of only 1\,\% would give the neutron star 
a recoil velocity of a few hundred kilometers per second;
\item The ejection of massive X-ray binaries from the clusters due to close 3- and 4-body encounters 
\citep{Poveda67,Heggie75,Mikkola83b,Moeckel10,Kroupa10,Mapelli11}.
\end{enumerate}

The first two mechanisms cannot explain the data, because the binary is ejected after the SN explosion,
and the system stays in the cluster for all the period of the primary evolution. 
This time is $\approx 2.5$~Myr for $\approx 100\msun$ star \citep{Starburst99},
which leaves no time for relocation to their present positions.

Alternatively, the ejection of massive X-ray binaries from the clusters can occur already 
at the very beginning stages of the cluster evolution due to close encounters. 
This scenario is supported by the presence of the blue halos of massive stars around the clusters \citep{Whitmore99}
as seen in Fig.~\ref{fig:ulx6}. The belts of young blue stars close to the young clusters might indicate that some 
of the less massive clusters have already dissolved \citep{Moeckel10}. 
Due to gas expulsion, young embedded star clusters become super-virial and start to expand
on the typical time-scale of $\approx 1$~Myr \citep{Kroupa01,PAK06}. 
The relatively large displacement and high spatial velocities are consistent with the results of 
N-body simulations showing that the most massive stars (binaries) are ejected at early stages of formation of stellar 
clusters \citep{Moeckel10,Mapelli11}.

The short cluster life-times imply limitation on the ULX binary mass ratio. 
One may understand  the association of the five ULXs with  young clusters (see Table~\ref{tab:clusters}),
if both companions are very massive $M_1 \sim M_2 \sim 100\msun$ 
and they evolve at approximately the same time-scale $T_1\approx 3$~Myr. 
If the original mass ratio  is large $M_1\sim 100\msun \gg M_2$, 
the evolution of the pre-ULX system takes twice as long time $\sim 2 T_1$, 
because after time $T_1$ the primary transfers its mass to the companion, 
which then evolves at the time-scale of $\sim T_1$.\footnote{We note that the companion has to be evolved
to produce a super-Eddington accretion rate required to power a ULX \citep{K01,K04}.}
Depending on the original mass ratio (and the masses) the total binary evolution takes between $T_1$ and $\sim 2 T_1$. 
Two sources,  X-11 and X-35, may be associated with less massive binaries 
with $M_1\sim$50--100$\msun$ and $M_2 \sim 50\msun$. 

The observations, however, are clearly inconsistent with the IMBH scenario as such massive black hole cannot be 
ejected from clusters at the required high ($\sim$80~km~s$^{-1}$) velocities. In their displacements from young
clusters, the ULXs do not differ from the sub-ULXs sources. In this property, the ULXs constitute  a continuation 
of the high-mass X-ray binary population.

\section{Conclusions}

In this paper we have studied the environments around the brightest X-ray sources in the Antennae galaxies. 
Using a high-accuracy astrometric solution we found a highly significant association between the X-ray sources and  the stellar clusters. 
We also showed that most of the bright X-ray sources are located outside of these clusters confirming previous findings. 
We have studied two samples of the X-ray sources, the ULX-sample ($L_{\rm X} \gtrsim 2\times 10^{39}$~erg~s$^{-1}$) 
and sub-ULX one ($3\times10^{38} \lesssim L_{\rm X} \lesssim 2\times 10^{39}$~erg~s$^{-1}$).
Using VIMOS imaging we were able to determine the reddening towards to the clusters near the ULXs and to find their age, 
which was  less than 5~Myr. 
The sub-ULX sample sources are also connected with young clusters ($\lesssim 10$~Myr) because they surrounded by
H$\alpha$ emission. 

The clear association of the young star clusters and the X-ray sources both in the ULX and sub-ULX samples in the Antennae 
demands these massive binaries to be ejected off the star clusters. All these sources belong to the runaway star population.
The extremely young ages of the clusters associated with the ULXs put the lower limit on the mass of 
their progenitor stars. The minimum possible mass varies between 40 and 100$\msun$, indicating that all 
studied ULXs in the Antennae galaxies are associated with the most massive binary stars. 
A few ULX sources associated with the youngest star clusters must originate from $\sim 100\msun$ 
stars with the mass ratio $\sim 1$. 
We conclude that the association with young star clusters and the displacements are the common properties of all 
high-mass X-ray binaries and the ULXs do not differ in this aspect from the less bright X-ray binaries.
Our findings strongly rule out an alternative hypothesis that majority of ULXs are IMBHs.

The star ejection mechanisms based on a SN explosion (either symmetric explosion in a close binary or a direct 
``kick'', imparted through asymmetry of the SN) are ruled out because the binary is ejected after the SN explosion.
Because the ULX-clusters are very young, 3--4~Myr, there is no time for the pre-SN evolution inside the cluster. 

The relatively large displacements (up to 300~pc) and high spatial velocities ($\sim$80~km s$^{-1}$) found
are consistent with the results of N-body simulations showing that the most massive stars (binaries) are 
ejected at the very beginning stages of the cluster evolution due to close encounters.

%%%%%%%%%%%%%%%%%%%%%%%%%%%%%%%%%%%%%%%%%%%%%%%%%%%%%%%%%%%%

\section*{Acknowledgments}
This work is based on observations made with the NASA/ESA Hubble Space Telescope, 
obtained from the data archive at the Space Telescope Science Institute. 
STScI is operated by the Association of Universities for Research in Astronomy, Inc. under NASA contract NAS 5-26555. 
This research used the facilities of the Canadian Astronomy Data Centre  
operated by the National Research Council of Canada with the support of the Canadian Space Agency. 
The research was supported by the Academy of Finland grants 127512 and 133179,
the Russian RFBR grant 10-02-00463 and Russian grant of the Scientific Leading Schools 4308.2012.2.  
We thank an anonymous referee for many useful comments. We also thank Thomas Tauris for discussion.

%\bibliographystyle{mn2e}
%\bibliography{allbib}
%\label{lastpage}
%\end{document}

\appendix

\section{VIMOS imaging, spectra of clusters  and comments on individual sources}
 
%%%%
\begin{figure*}
\includegraphics[width=0.95\linewidth]{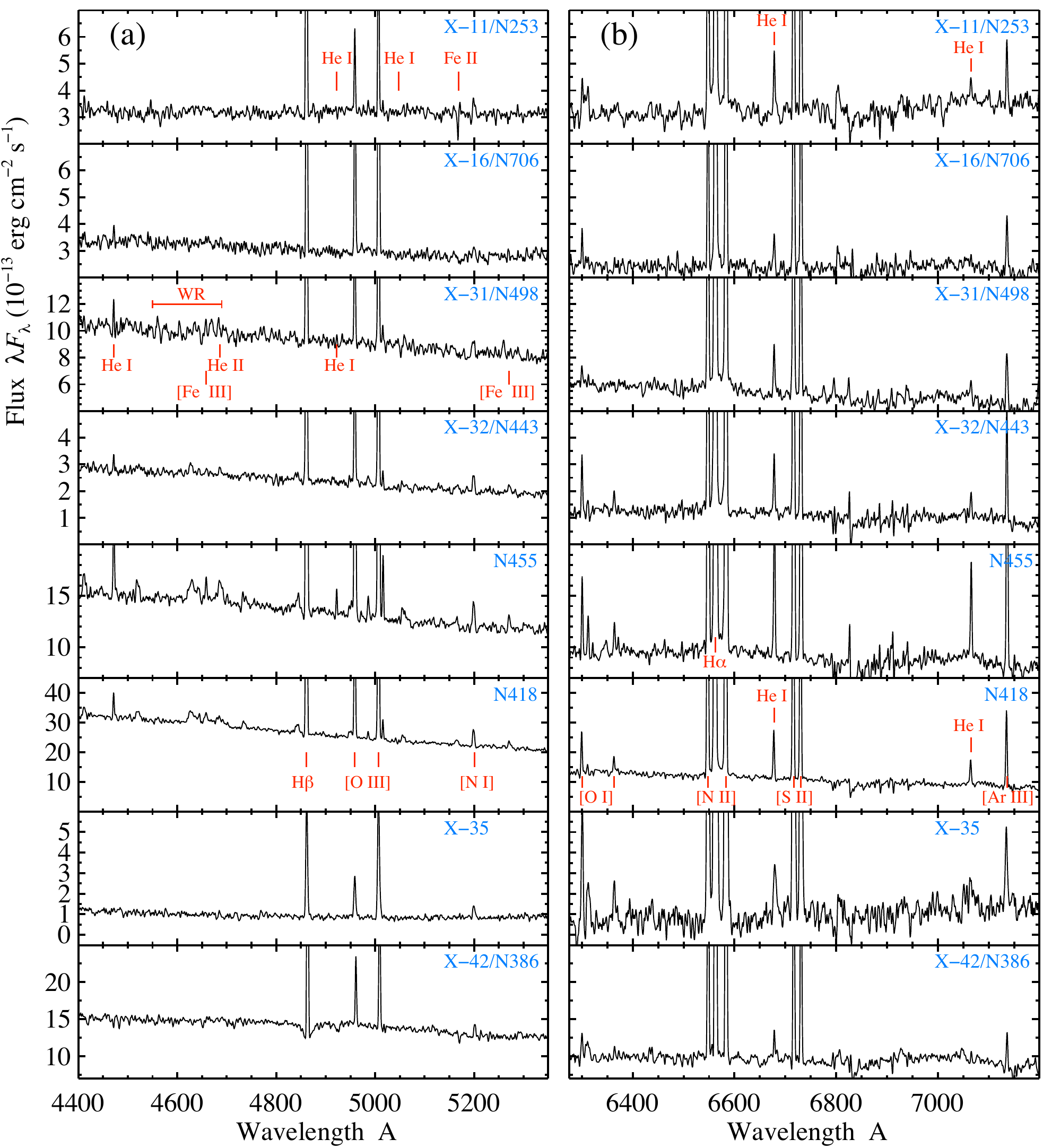}
\caption{The VIMOS spectra of the clusters associated with X-11, 16,  31, 32, 35 and 42  
(see Table~\ref{tab:clusters})
as well as the cluster complexes N455 and N418 (objects 5 and 6 in 
\citealt{Bastian06}, see Fig.~\ref{fig:6mapsX32}) 
taken with the blue (panel a) and orange (panel b) grisms. 
The main spectral lines are indicated. 
}
\label{fig:spectra}
\end{figure*}
%%%%

\begin{figure*}
\includegraphics[width=12cm]{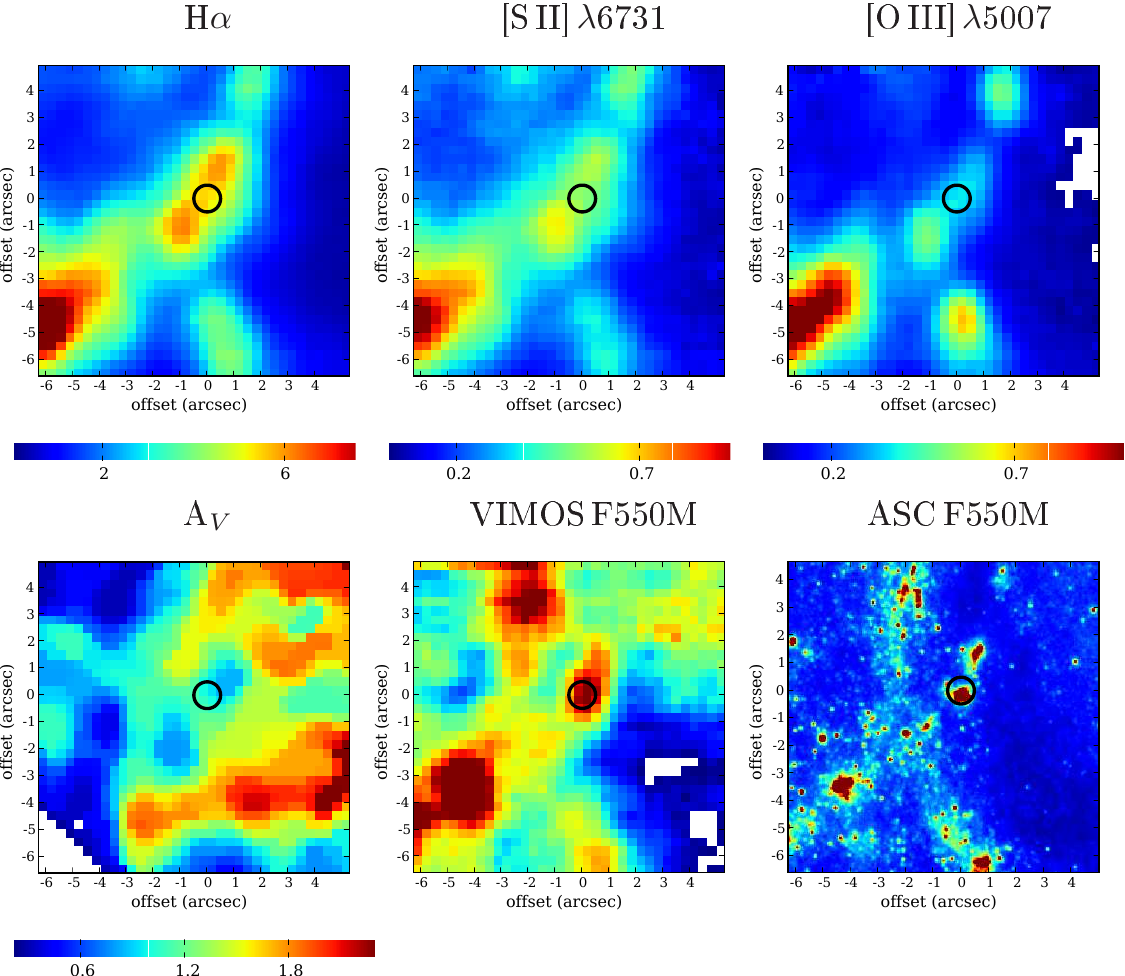}
\caption{VIMOS images of the region around X-11 with the X-ray position marked by circles (with 0\farcs5 radius). 
The images show the line intensities in H$\alpha$, [S\,{\sc ii}]~$\lambda$6731, [O\,{\sc iii}]~$\lambda$5007, 
the extinction $A_V$, the continuum flux measured by VIMOS in the ACS F550M band and 
the ACS F550M continuum image itself. 
The line intensities  (measured in $10^{-16}$~erg cm$^{-2}$ s$^{-1}$ pixel$^{-1}$) were not dereddened.  
North is to the right and east is up. Here 1\arcsec=107 pc.
}
\label{fig:6mapsX11}
\end{figure*}

\begin{figure*}
\includegraphics[width=12cm]{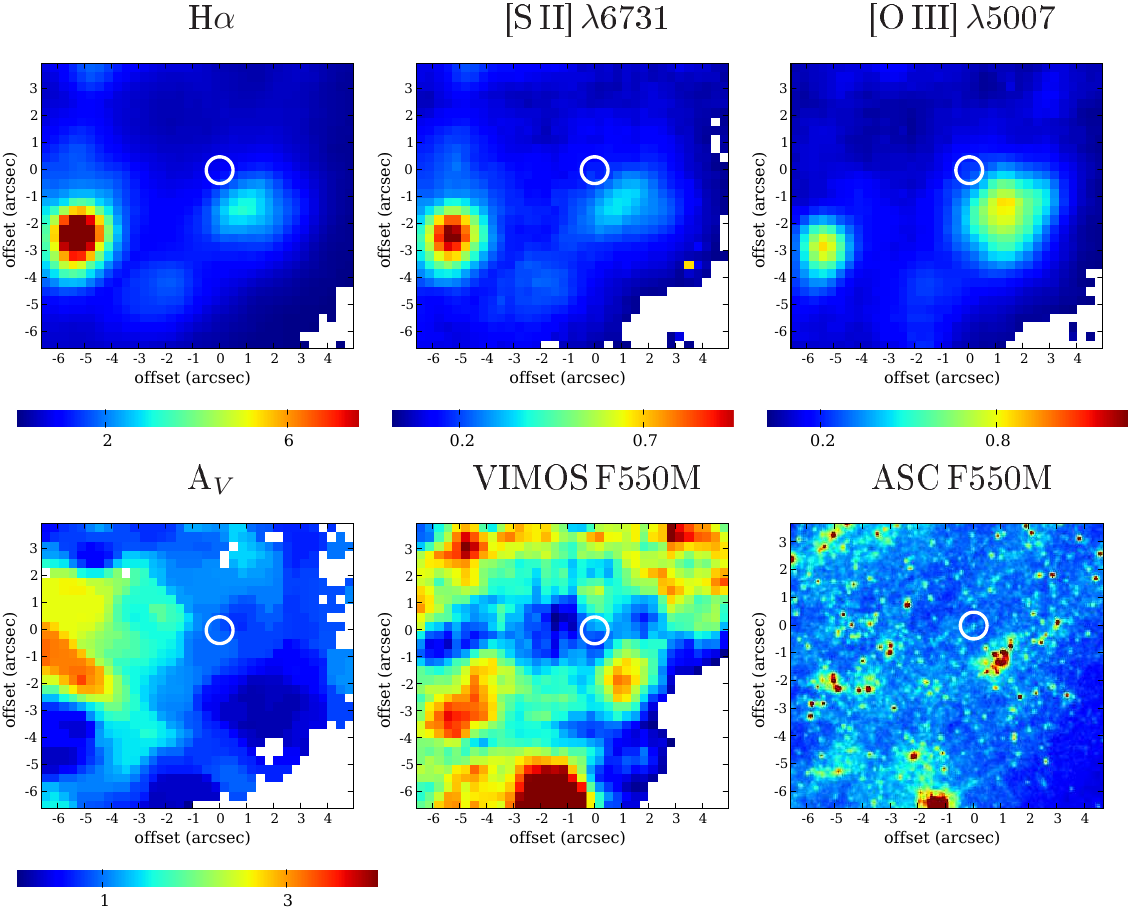}
\caption{Same as  Fig.~\ref{fig:6mapsX11}, but for the source X-16. 
}
\label{fig:6mapsX16}
\end{figure*}

\begin{figure*}
\includegraphics[width=12cm]{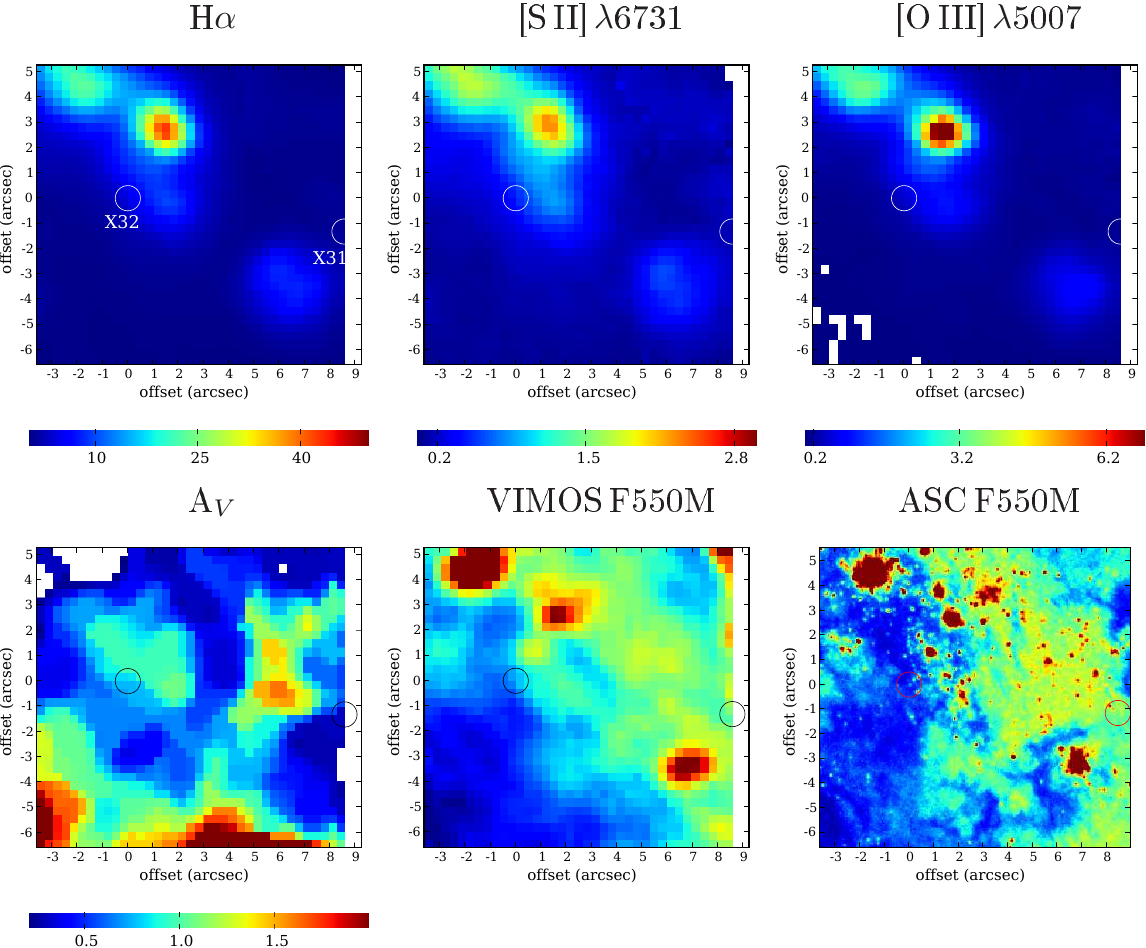}
\caption{Same as  Fig. \ref{fig:6mapsX11}, but for the sources X-31 and X-32. 
Two stellar complexes N455 and N418  (objects 5 and 6 from the study of \citealt{Bastian06}) are 
located $\sim$3\farcs5 to the NE and $\sim$5\arcsec  to the SE from the X-ray positions of X-32, respectively. 
The nearest clusters to X-31 and X-32 are situated $\sim$3\arcsec SW  and $\sim$1\arcsec NE  from the corresponding source.  
}
\label{fig:6mapsX32}
\end{figure*}

\begin{figure*}
\includegraphics[width=13cm]{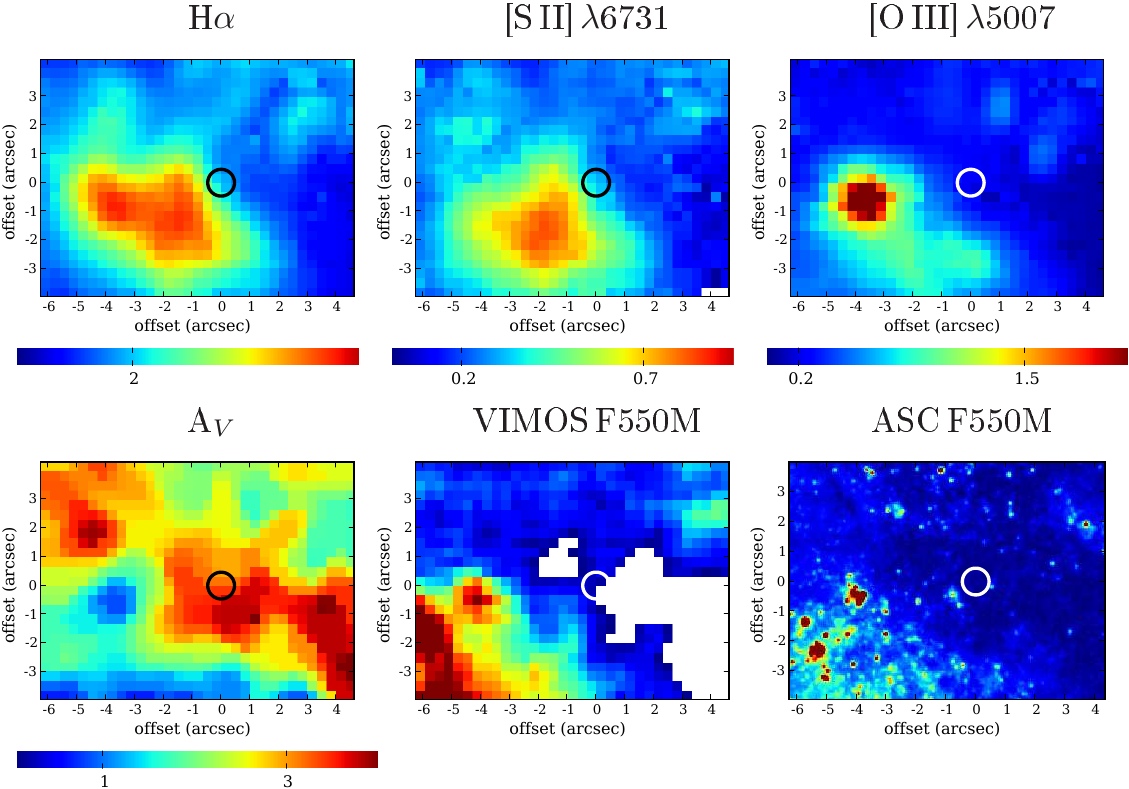}
\caption{Same as  Fig.~\ref{fig:6mapsX11}, but for the source X-35. 
}
\label{fig:6mapsX35}
\end{figure*}

\begin{figure*}
\includegraphics[width=13cm]{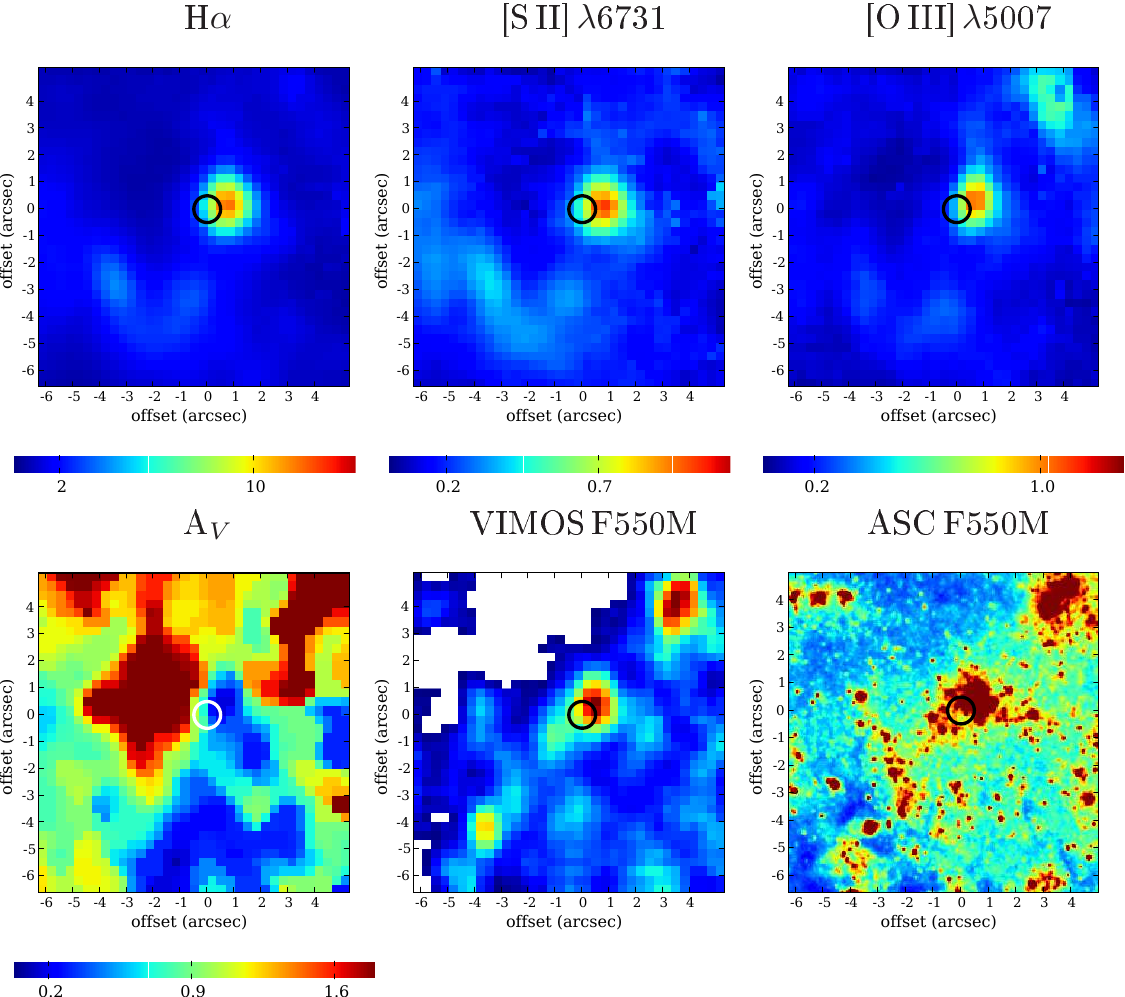}
\caption{Same as  Fig.~\ref{fig:6mapsX11}, but for the source X-42. 
\label{lastpage}
}
\label{fig:6mapsX42}
\end{figure*}

Fig.~\ref{fig:spectra} presents the spectra of the bright stellar clusters next to the 
selected X-ray sources. 
In Figs~\ref{fig:6mapsX11}--\ref{fig:6mapsX42} we show VIMOS images of regions around six 
brightest X-ray sources.
The images were derived in H$\alpha$, [S\,{\sc ii}]~$\lambda$6731, [O\,{\sc iii}]~$\lambda$5007. 
There are also extinction $A_V$ maps and images in the continuum extracted in the ACS/F550M band and the 
ACS F550M image itself for comparison. In the estimates of the reddening we used the H$\alpha$/H$\beta$ flux 
ratios measured from the  nebulae emission in the VIMOS fields. 
We adopt $R_V=3.1$ in the estimates. 

%%%%

%%%%
 
The source X-18 (see Fig.~\ref{fig:ulx6})
is located in a  dusty zone, where there are no clusters from WS95 catalogue. 
There are also no known clusters around the source X-19. 
Source X-35 is located in a dust lane with two clusters at about the same distance (0\farcs6--0\farcs8) nearby. 
We have studied (Table~\ref{tab:clusters})  the brighter object, which is not from the WS95 catalogue, 
because another cluster is too weak for Êa detailed study.  
The objects associated with the X-ray sources X-16 and X-42 are cluster complexes and 
the brightest cluster in the complex was selected.

\subsection{X-11}

This X-ray source is located inside a cluster with dereddened $V=18.8$ 
(see Figs~\ref{fig:ulx6} and \ref{fig:6mapsX11}, Table~\ref{tab:sources}).
The cluster's spectrum (see Fig.~\ref{fig:spectra}) has strong emission lines, 
which are formed in H\,{\sc ii} regions, however, H$\alpha$
line has broad wings indicating their stellar origin. 
In addition to the hydrogen lines H$\alpha$ and H$\beta$
we see  lines  [O\,{\sc i}]~$\lambda \lambda$6300, 6363, [O\,{\sc iii}]~$\lambda
\lambda$4959, 5007, [S\,{\sc ii}]~$\lambda \lambda$6716, 6730, [N\,{\sc ii}]~$\lambda\lambda$6548, 6583, 
[N\,{\sc ii}]~$\lambda$5754. 
There are also lines indicating strong UV ionization sources, they are He\,{\sc i}~$\lambda \lambda$4922, 5015, 5876,
6668, 7065, [N\,{\sc i}]~$\lambda \lambda$5197, 5200 and [Ar\,{\sc iii}]~$\lambda$7135.
Although the [O\,{\sc i}] lines are not resolved, other nebular lines are slightly broadened. 
Correcting for the spectral resolution we find FWHMs of hydrogen, [O\,{\sc iii}], [Ar\,{\sc iii}] lines of 50, 70 and 80\,km s$^{-1}$ respectively. 
There are numerous Fe\,{\sc ii} narrow lines both in emission and in absorption and 
absorption line of Mg\,{\sc ii}~$\lambda$4481. The strongest Fe\,{\sc ii}~$\lambda$5169
line shows clear P Cygni-like profile. The absorption lines indicate their
origin in stellar atmospheres. In the red spectral region, the Fe\,{\sc ii} lines appear in emission. 
We suspect a presence of the [Fe\,{\sc ii}], N\,{\sc ii}, N\,{\sc iii} and S\,{\sc ii} emission lines in the spectrum, 
however, the spectrum quality is not enough to study these lines.

\subsection{X-16}

This X-ray source is located close (1\farcs5) to a  cluster with dereddened $V=19.71$  
(see Figs~\ref{fig:ulx6} and~\ref{fig:6mapsX16}, Table~\ref{tab:sources}). 
The nebular lines in the cluster spectrum are about the same as those in cluster next to X-11, 
but the [O\,{\sc iii}]~$\lambda$5007 line is brighter that H$\beta$, which
indicates a stronger UV ionization source. The narrow H$\alpha$ shows broad wings. 
The Fe\,{\sc ii} absorption lines and weak Si\,{\sc ii}~$\lambda \lambda$6347, 6371 lines are present. 
 
\subsection{X-31 and X32}

The spectra of the clusters located close to the X-ray sources X-31 (N498, dereddened $V=18.5$,
separation 2\farcs5) and X-32 (N443, $V=18.6$, separation 1\farcs5) are shown in Fig.~\ref{fig:spectra}
together with the spectra of two other clusters N455 and N418 
(named 5 and 6 in \citealt{Bastian06}). 
Cluster N455 is located  $\sim$ 3\farcs5 to the NE and the brigth cluster N418 is located $\sim$ 5\arcsec  
to SE from X-ray positions of X-32 (see Figs~\ref{fig:wholemap}, \ref{fig:ulx6} and \ref{fig:6mapsX32}). 
We show spectra of these four clusters together because they are similar and all clusters are located close to each other. 
N455 and N418 show bright Wolf-Rayet spectral features, He\,{\sc ii}~$\lambda$4686,
Bowen blend (C\,{\sc iii}/N\,{\sc iii} emission lines, where N\,{\sc iii} lines dominate strongly)
together with the numerous [Fe\,{\sc iii}] and [Fe\,{\sc ii}] emission lines. 
Emission lines of Fe\,{\sc ii} and Si\,{\sc iii} are also detected. There are N\,{\sc ii}--N\,{\sc iv}
emission lines as well, however, the Wolf-Rayet  hump at $\sim 5800${\AA} (C\,{\sc iv} and N\,{\sc iv}
lines) is not strong. The H$\alpha$ lines show broad emission wings in the spectra.
All these features point towards WN and LBV-like stars
with extended winds in N455 and N418 clusters, indicating their ages 
to be less than 6~Myrs. The spectra of the clusters, which we accosiate with X-31
and X-32, show similar features. The cluster next to X-31 shows H$\alpha$
line with broad wings, the N\,{\sc iii} lines in the Bowen blend, many [Fe\,{\sc iii}], [Fe\,{\sc ii}] and
N\,{\sc ii} emission lines, as well as [Ar\,{\sc iv}]~$\lambda \lambda$7170, 7237, 7262, 7331 lines
in addition to the bright [Ar\,{\sc iii}]~$\lambda$7135 line. Broad He\,{\sc ii}~$\lambda$4686 is
marginally detected. The cluster next to X-32 shows similar features, however, its [Fe\,{\sc iii}] emission lines 
are notably broadened, and no [Ar\,{\sc iv}] lines were detected in the spectrum. Emission lines of
Fe\,{\sc ii}, N\,{\sc ii} and Si\,{\sc iii} together with a faint He\,{\sc ii}~$\lambda$4686 hump are detected.

\subsection{X-35}

This X-ray source is located close (0\farcs6--0\farcs8) to two clusters. 
The closest cluster N115 from WS95 catalogue (0\farcs57 to the north)  
is too weak for Êa detailed study.  
Another  apparently rather faint cluster (not mentioned in WS95, with dereddened $V=18.8$; 
see Figs~\ref{fig:ulx6} and~\ref{fig:6mapsX35}, Table \ref{tab:sources}) 
is situated 0\farcs83 to the SSW from X-35 at  the edge of a strong dust line. 
The cluster is notably reddened ($A_V\sim3.5$) and its spectrum is noisy. 
We find about the same nebular spectrum as that in X-16 with the [O\,{\sc iii}]~$\lambda$5007 line 
being brighter that H$\beta$.
However, in the red spectral region He\,{\sc i} nebular lines and [Ar\,{\sc iii}] are strong and broadened.
Interestingly, there are [Fe\,{\sc iii}] emission lines in the spectrum like those we
observe in N\,5 and N\,6  clusters close to X-32. The [Fe\,{\sc ii}] emission lines are marginally detected.

\subsection{X-42}

This X-ray source in located close (at 0\farcs8) to a bright  cluster
complex with total (dereddened) $V=17.2$ 
(see Figs~\ref{fig:ulx6} and~\ref{fig:6mapsX42}, Tables~\ref{tab:sources} and \ref{tab:clusters}). 
The set of nebular lines is the same as in other clusters. 
However, this complex includes stars of different ages, there is a notable amount of A-type stars,
because the H$\beta$ line has broad absorption wings. There are many stellar
absorption lines in the spectrum indicating the B/A populations, numerous He\,{\sc i},  
Fe\,{\sc ii}, C\,{\sc ii} lines as well as the Mg\,{\sc ii} and Si\,{\sc ii} lines. He\,{\sc i} 
lines appear in emission in the red part of the spectrum.

\end{document}